# Depth-Resolved Thermal Conductivity of HFCVD Diamond Films via Square-Pulsed Thermometry

Kexin Zhang[1], Xiaosong Han[2,3], Ershuai Yin[4], Xin Qian[1], Junjun Wei[2,*], Puqing Jiang[1,*]

[1]*School of Energy and Power Engineering, Huazhong University of Science and Technology, Wuhan, Hubei 430074, China*

[2]*Institute for Advanced Materials and Technology, University of Science and Technology Beijing, Beijing, 100083, China*

[3]*Shunde Graduate School, University of Science and Technology Beijing, Foshan, Guangdong 528399, China*

[4]*School of Energy and Power Engineering, Nanjing University of Science and Technology, Nanjing, Jiangsu 210094, China*

**Abstract**

The integration of high-thermal-conductivity diamond films onto silicon carbide (SiC) substrates offers a promising pathway for thermal management in high-power electronic devices. Here, we investigate the depth-dependent thermal conductivity of a ~5 $\mu$m-thick diamond film grown on SiC by hot-filament chemical vapor deposition (HFCVD) using square-pulsed source (SPS) thermometry. Electron backscatter diffraction (EBSD) and transmission electron microscopy (TEM) reveal pronounced grain coarsening from the nucleation interface to the film surface. By combining frequency-dependent thermal penetration with a depth-resolved thermal transport model, we quantitatively reconstruct the thermal conductivity profile. The thermal conductivity increases sharply from ~60 W m$^{-1}$ K$^{-1}$ near the nucleation region to ~200 W m$^{-1}$ K$^{-1}$ at the surface, directly reflecting the underlying microstructural evolution. These results provide a physically grounded understanding of graded heat transport in HFCVD diamond and offer practical guidance for engineering diamond-based thermal management layers for next-generation power devices.

*Corresponding Authors: (J.W.) weijj@ustb.edu.cn; (P.J.) jpq2021@hust.edu.cn

The thermal management of high-power electronic devices has become increasingly critical as power densities continue to rise.[1] Silicon carbide (SiC) is a cornerstone material for such applications, yet its own heat dissipation capacity is often overwhelmed, leading to performance degradation and reliability concerns.[2, 3] Diamond, with the highest known room-temperature thermal conductivity ($\sim$2000 W m$^{-1}$ K$^{-1}$), represents a compelling solution for integrated heat spreading. However, the practical integration of high-quality diamond with SiC remains challenging. Although the lattice mismatch between diamond and common hexagonal SiC polytypes is relatively small, the large disparity in thermal expansion coefficients generates substantial compressive stress during cooling from the growth temperature. As a result, directly grown diamond films commonly suffer from cracking, poor interfacial adhesion, and severe limitations in achievable thickness.[4-6] These constraints motivate the exploration of deposition techniques that balance film quality, scalability, and cost.

While microwave plasma chemical vapor deposition (MPCVD) produces the highest quality diamond films, its high cost and limitations in scaling for large-area substrates constrain its widespread industrial adoption.[7] In this context, hot filament chemical vapor deposition (HFCVD) emerges as a compelling, technologically viable alternative. HFCVD offers superior scalability, lower operational costs, and compatibility with non-planar geometries, making it a more practical route for implementing diamond thermal management in commercial devices.[8] The central challenge, therefore, shifts from merely growing diamond to engineering a cost-effective HFCVD diamond film with optimized thermal performance despite its inherent microstructural imperfections.

A key, yet often overlooked, aspect of polycrystalline diamond films is their pronounced structural gradation. HFCVD films typically nucleate with fine, defective crystallites and impurities, evolving towards larger, more thermally conductive grains at the surface.[9, 10] This intrinsic heterogeneity implies that the thermal conductivity is not a uniform bulk property but is depth-dependent.[11-13] A conventional, bulk-averaged thermal measurement obscures this critical detail and fails to provide the necessary insight for targeted process optimization. Understanding the depth-resolved thermal transport is paramount for answering fundamental questions: How quickly does the thermal conductivity improve from the defective nucleation region? What is the ultimate thermal performance of the higher-quality surface layer? The answers are essential for making informed trade-offs between film thickness, growth time, cost, and final thermal performance.

In this work, we address this critical knowledge gap by employing a sophisticated square-pulsed source (SPS) thermometry technique to directly measure the depth-dependent thermal conductivity of HFCVD diamond films grown on SiC. We correlate the thermal transport profile with meticulous microstructural characterization from electron backscatter diffraction (EBSD) and transmission electron microscopy (TEM). By fitting the frequency-dependent thermal data to a depth-resolved conductivity model, we quantitatively reconstruct the thermal conductivity as a function of film depth. Our findings reveal a significant gradient in thermal conduction, directly linking it to the microstructural evolution. This study provides not only a fundamental understanding of heat transport in graded polycrystalline films but also a practical framework for guiding the development of high-performance, cost-effective diamond thermal solutions for next-generation electronics.

The experimental details are briefly summarized as follows. Polycrystalline diamond films were deposited on 4H-SiC substrates using HFCVD. An optimized filament-substrate distance of 8 mm and a filament power of 900 W per tantalum filament were employed to minimize intrinsic stress and maximize crystalline quality. The optimization process, including SEM and Raman studies of the diamond films grown using different filament-substrate distances and filament powers, is described in detail in Supplementary Information Sec. S1. Stable filament carbonization was ensured through precise current control and a 1-hour pre-treatment step, producing a consistent gas-phase environment for diamond nucleation.

To enhance diamond nucleation density and mitigate thermal expansion mismatch during high-temperature growth, a ~20 nm amorphous silicon nitride (SiN) interlayer was introduced as a compliant buffer layer between the diamond film and the SiC substrate. The SiN layer was deposited using magnetron sputtering at 200°C and 1 Pa, conditions determined to yield a high N/Si ratio, smooth morphology, and effective diffusion-barrier behavior. The resulting multilayer stack consists of a 5.27 μm diamond film on a SiN/SiC substrate.

Cross-sectional TEM imaging (Figs. 1(a,b)) reveals a distinct dark-contrast nucleation transition zone at the interface. EBSD band-contrast and phase maps (Figs. 1(c,d)) further show a pronounced grain-size gradient along the growth direction: nanoscale or partially amorphous crystallites near the interlayer evolve into well-faceted diamond grains toward the top surface.

During high-temperature growth, the SiN interlayer undergoes partial pyrolysis, releasing nitrogen that desorbs as $N_2$ and enabling silicon diffusion into the carbon-rich growth front. The diffusing Si forms a transient SiC-rich transition layer, which diminishes as carbon supersaturation increases and diamond growth proceeds. As evidenced in Fig. 1(d), this process leads to a broader transition region than the nominal SiN thickness and produces a smooth, compositionally graded boundary between SiC and diamond. Despite the diffusion-broadened zone, the overlying diamond remains well-crystallized, and the interface exhibits excellent mechanical stability, which is essential for reliable thermal and structural performance.

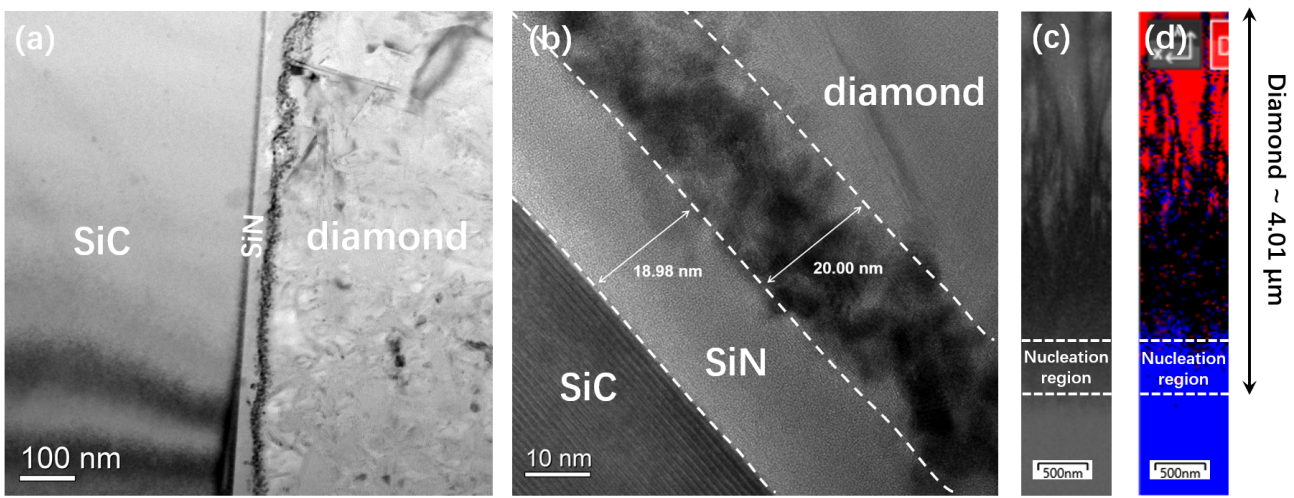

**Figure 1.** Cross-sectional microstructure of the HFCVD diamond/SiN/SiC stack. (a,b) TEM images showing the ~20 nm SiN interlayer and the adjacent nucleation transition zone. (c,d) EBSD band-contrast and phase maps revealing pronounced grain-size coarsening from the fine-grained nucleation region to the faceted top surface. These microstructural gradients underpin the depth dependence of thermal conductivity in the diamond film.

The depth-dependent thermal conductivity of the diamond film was measured using square-pulsed source (SPS) thermoreflectance.[14, 15] In this technique, a square-wave modulated pump laser induces periodic heating, and the resulting surface temperature oscillation amplitude is directly measured in the time domain using a waveform analyzer. Because the analysis relies on amplitude detection rather than phase tracking, SPS avoids phase-drifting artifacts inherent to phase-sensitive methods and provides improved measurement robustness and signal-to-noise ratio.[16]

A defining characteristic of SPS thermoreflectance is its broad modulation-frequency range, extending from a few hertz to tens of megahertz. The modulation frequency $f$ controls the thermal penetration depth $d_p$, such that higher frequencies probe near-surface regions while lower frequencies access deeper layers. When $d_p$ is smaller than the diamond film thickness $h_{\text{dia}}$, the measured effective thermal conductivity $k_{\text{eff}}(f)$ corresponds to a depth-weighted average of the thermal transport properties within the thermally sampled region. Owing to this frequency-domain selectivity, SPS thermoreflectance has been widely applied to investigate thermal transport in a broad range of materials and multilayered structures.[17-23]

In HFCVD diamond films, grain coarsening during growth produces a fine-grained, defect-rich nucleation layer beneath a coarse-grained surface region. As a result, phonon transport is primarily influenced by grain-boundary scattering, leading to a cross-plane thermal conductivity that is highest near the growth surface and decreases toward the substrate. To capture this depth dependence, we describe the local thermal conductivity as

$$k(z) = k_1 + (k_0 - k_1)e^{-\alpha z}, \qquad (1)$$

where $k_0$ and $k_1$ denote the thermal conductivities at the surface and nucleation region, respectively, $\alpha$ is an attenuation coefficient, and $z$ denotes the depth measured from the film surface along the cross-plane direction.

The SPS measurement, however, yields an effective thermal conductivity $k_{\text{eff}}(f)$, which represents an average over a frequency-dependent probing depth. An intuitive approximation assumes that thermal waves probe uniformly down to a thermal penetration depth,

$$d_p(f) = \sqrt{\frac{k_{\text{eff}}(f)}{\pi f C}}, \tag{2}$$

where $C$ is the volumetric heat capacity. Within this picture, $k_{\text{eff}}(f)$ corresponds to the harmonic average of $k(z)$ over $0 \leq z \leq d_p$:

$$k_{\text{eff}}^{-1}(f) = \frac{1}{d_p(f)} \int_0^{d_p(f)} k^{-1}(z) \, dz. \tag{3}$$

While this model correctly reproduces the bulk limit at low frequencies ($d_p \gg h_{\text{dia}}$) and surface limit at high frequencies ($d_p \to 0$), it exhibits systematic deviations at intermediate frequencies. The sharp cutoff at $z = d_p$ neglects both the exponential decay and phase interference of thermal waves in gradient media, which becomes important when the thermal conductivity varies on length scales comparable to the thermal diffusion length.

A more physically consistent description expresses the inverse effective conductivity as a depth-weighted average:

$$k_{\text{eff}}^{-1}(f) = \int_0^\infty k^{-1}(z) \, W(z, f) \, dz, \tag{4}$$

where $W(z, f)$ is a normalized weighting function satisfying $\int_0^\infty W(z, f) \, dz = 1$. For a homogenous material, $W(z, f)$ decays exponentially with depth. For a graded film with slowly varying $k(z)$, the Wentzel-Kramers-Brillouin (WKB) approximation gives[24]

$$W(z, f) \propto \exp\left[-2\sqrt{\pi f C} \int_0^z \frac{d\xi}{\sqrt{k(\xi)}}\right], \tag{5}$$

where the integral accounts for cumulative attenuation through the depth-dependent local thermal diffusion length $\mu(\xi) = \sqrt{k(\xi)/(\pi f C)}$. Normalizing $W(z, f)$ yields

$$k_{\text{eff}}^{-1}(f) = \frac{\int_0^\infty k^{-1}(z) \exp\left[-2\sqrt{\pi f C} \int_0^z \frac{d\xi}{\sqrt{k(\xi)}}\right] dz}{\int_0^\infty \exp\left[-2\sqrt{\pi f C} \int_0^z \frac{d\xi}{\sqrt{k(\xi)}}\right] dz}. \tag{6}$$

This formulation naturally reduces to the harmonic-average result for uniform $k(z)$, while more generally accounting for the exponential attenuation of thermal waves in non-uniform media. The detailed derivation of this weighted harmonic mean model is provided in Supplementary Information, Sec. S2.

The thermal model, illustrated in Fig. 2, thus consists of a self-consistent set of Eqs. (1) and (6). For each measurement frequency, the equations are solved numerically to obtain the corresponding effective thermal conductivity. The depth-dependent conductivity profile is then reconstructed by fitting the parameters $k_0$, $k_1$, and $\alpha$ in Eq. (1), using a least-squares minimization of the difference between the calculated and measured $k_{\text{eff}}(f)$ across all frequencies. Details of the numerical implementation, including the solution of the self-consistent equations and the parameter optimization procedure, are provided as MATLAB code in Supplementary Information (Sec. SA).

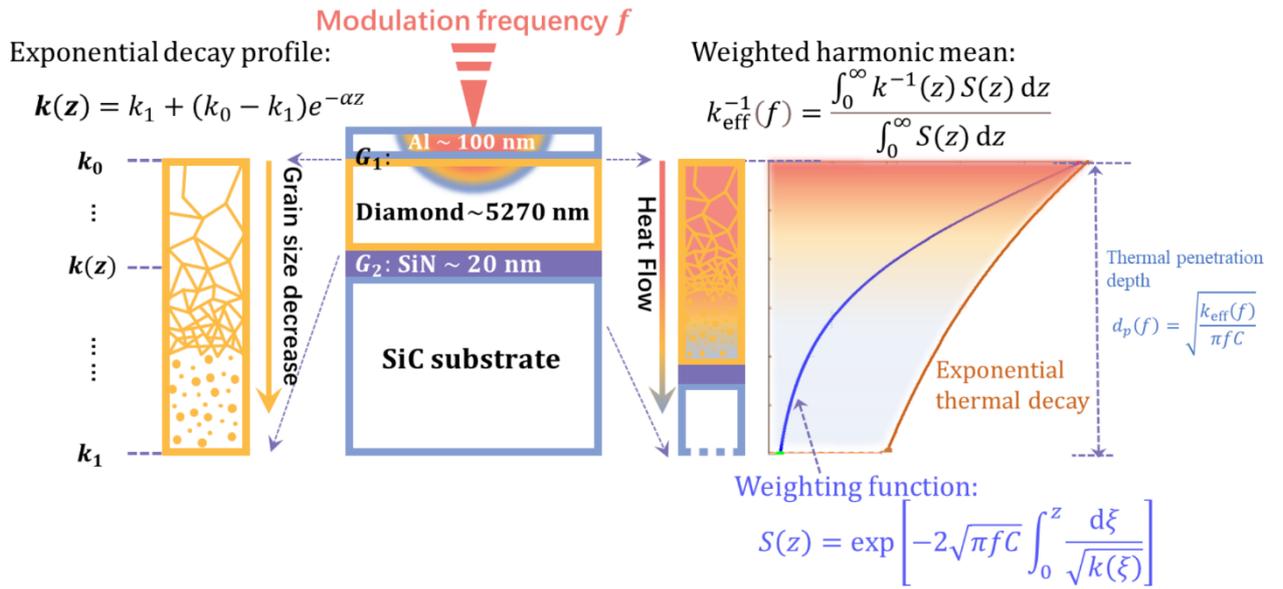

**Figure 2.** Schematic of the thermal model used to extract the depth-resolved thermal conductivity. The modulation frequency controls the thermal penetration depth, enabling frequency-dependent probing of different depths in the diamond film. The measured effective conductivity at each frequency $k_{\text{eff}}(f)$ is linked to the intrinsic profile $k(z)$ through a weighted harmonic mean relation.

Depth-resolved probing of the diamond film was achieved by acquiring SPS signals at eight modulation frequencies, ranging from 15 MHz to 100 kHz. Representative fitted waveforms and sensitivity curves for 10 MHz, 500 kHz, and 100 kHz are displayed in Fig. 3. Additional fitted signals and sensitivity curves for all other modulation frequencies can be found in the Supplementary Information Sec. S4. In the multilayer thermal model, the free fitting parameters were the cross-plane thermal conductivity of diamond $k_{z,\text{dia}}(f)$ (with a distinct value fitted at each of the eight frequencies), its in-plane thermal conductivity $k_{r,\text{dia}}$, and the two interfacial thermal conductances ($G_1$ for Al/diamond and $G_2$ for diamond/SiC). This resulted in a total of 11 fitted parameters. All other parameters, including the properties of the non-diamond layers ($k_{\text{Al}}$, $C_{\text{Al}}$, $h_{\text{Al}}$, $k_{r,\text{SiC}}$, $k_{z,\text{SiC}}$, $C_{\text{SiC}}$, $r_0$) and the diamond's heat capacity and thickness ($C_{\text{dia}}$, $h_{\text{dia}}$), were treated as known quantities, either experimentally measured or obtained from literature. The full lists of these input and fitted parameters, including their values, uncertainties, and determination methodology, are detailed in the Supplementary Information (Tables S1 and S2).

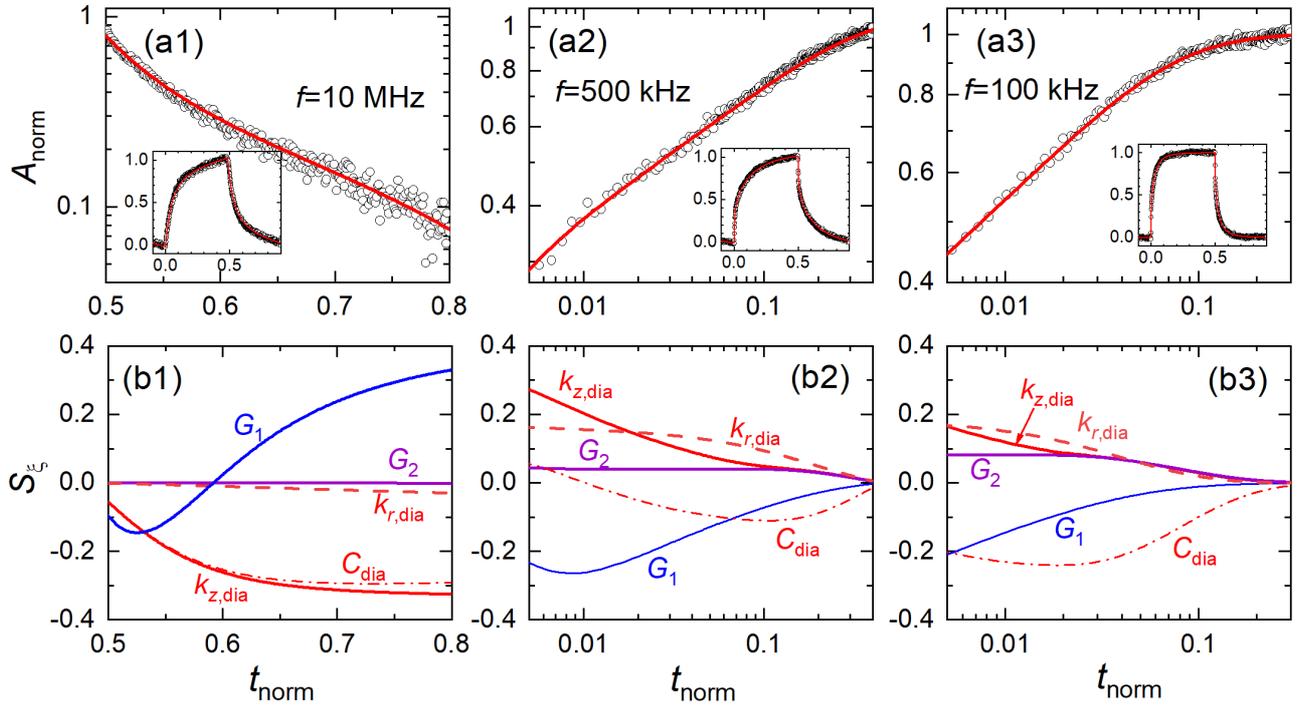

**Figure 3.** Representative SPS waveforms and sensitivity analyses at three modulation frequencies. (a1-a3) Measured signals (dots) and best-fit curves (lines). (b1-b3) Corresponding sensitivity coefficients showing how the probed depth evolves with frequency. High frequencies (e.g., 10 MHz) are dominated by the near-surface conductivity and the top interface, mid-range frequencies (500 kHz) reflect the bulk of the diamond layer, and low frequencies (100 kHz) become sensitive to the diamond/SiC interfacial conductance.

At 10 MHz, the thermal penetration depth is shallow, making the SPS response primarily sensitive to $k_{z,\text{dia}}$ and $G_1$, with negligible sensitivity to $k_{r,\text{dia}}$ or $G_2$ [Fig. 3(b1)]. Since $G_1$ mainly affects the cooling slope while $k_{z,\text{dia}}$ affects the signal amplitude, these two parameters are not coupled and can be extracted simultaneously. The resulting best-fit values are $k_{z,\text{dia}} = 130 \pm 14.6$ W m$^{-1}$K$^{-1}$ and $G_1 = 70 \pm 4$ MW m$^{-2}$ K$^{-1}$. This $G_1$ value was subsequently fixed for the analysis of all lower-frequency data.

When the frequency is reduced to 500 kHz, the thermal penetration depth extends deeper into the film, making the SPS signals sensitive to both $k_{z,\text{dia}}$ and $k_{r,\text{dia}}$ [Fig. 3(b2)]. The resulting fitted values are $k_{z,\text{dia}} = 88 \pm 8.8$ W m$^{-1}$K$^{-1}$ and $k_{r,\text{dia}} = 15 \pm 1.7$ W m$^{-1}$K$^{-1}$. The thermal penetration depth at this frequency is estimated as $d_p = 5.48\ \mu$m, which is comparable to the film thickness. This indicates that the extracted conductivity represents an effective value for nearly the entire diamond film. Furthermore, sensitivity to $G_2$ remains low, meaning its exact value has a negligible impact on the fit.

With a further decrease in frequency to 100 kHz, $d_p$ exceeds $h_{\text{dia}}$, shifting the sensitivity away from $k_{z,\text{dia}}$ and toward the substrate thermal boundary conductance $G_2$ [Fig. 3(b3)]. This progression of sensitivity with frequency enables the determination of $G_2$ through a simultaneous fit of the data from 500 kHz to 100 kHz, yielding a value of $G_2 = 50 \pm 14$ MW m$^{-2}$ K$^{-1}$.

The results indicate that for frequencies above 500 kHz, with the condition $d_p < h_{\text{dia}}$ being satisfied, the retrieved $k_{z,\text{dia}}(f)$ values represent the depth-averaged effective conductivity $k_{\text{eff}}(f)$ over the corresponding $d_p(f)$. This provides six experimental data pairs of $k_{\text{eff}}(f)$ versus $d_p(f)$, plotted as open circles in Fig. 4. In contrast, for frequencies below 500 kHz where $d_p > h_{\text{dia}}$, the retrieved $k_{z,\text{dia}}(f)$ becomes independent of $d_p(f)$, as shown in the inset of Fig. 4. This plateau indicates that the film thickness, rather than $d_p$, now limits the phonon mean free path.

The high-frequency data ($f > 500$ kHz) were fitted using the implicit depth-weighted relation in Eq. (6), which links the measured $k_{\text{eff}}(f)$ to the intrinsic conductivity profile $k(z)$. A least-squares fit (blue curve in Fig. 4) yields $k_0 = 205 \pm 19.5$ W m$^{-1}$ K$^{-1}$, $k_1 = 57.8 \pm 6.4$ W m$^{-1}$ K$^{-1}$, and $\alpha = 1.055 \pm 0.306\ \mu$m$^{-1}$. The reconstructed $k(z)$ profile (red solid curve in Fig. 4), generated from these parameters, increases monotonically from the fine-grained nucleation region toward the coarse-grained growth surface, consistent with the grain-size evolution observed by EBSD.

For comparison, the simplified harmonic-mean model (Eq. (3)) was also applied, yielding the reconstructed $k(z)$ profile shown by the pink dashed curve in Fig. 4. While both models produce qualitatively similar depth-dependent trends, the simplified model overestimates the nucleation-region conductivity $k_1$ by approximately 30% (from 58 to 74 W m$^{-1}$ K$^{-1}$). This discrepancy stems from the simplified model's inherent limitation: it does not consider the gradual attenuation and wave interference effects that occur within a graded film.

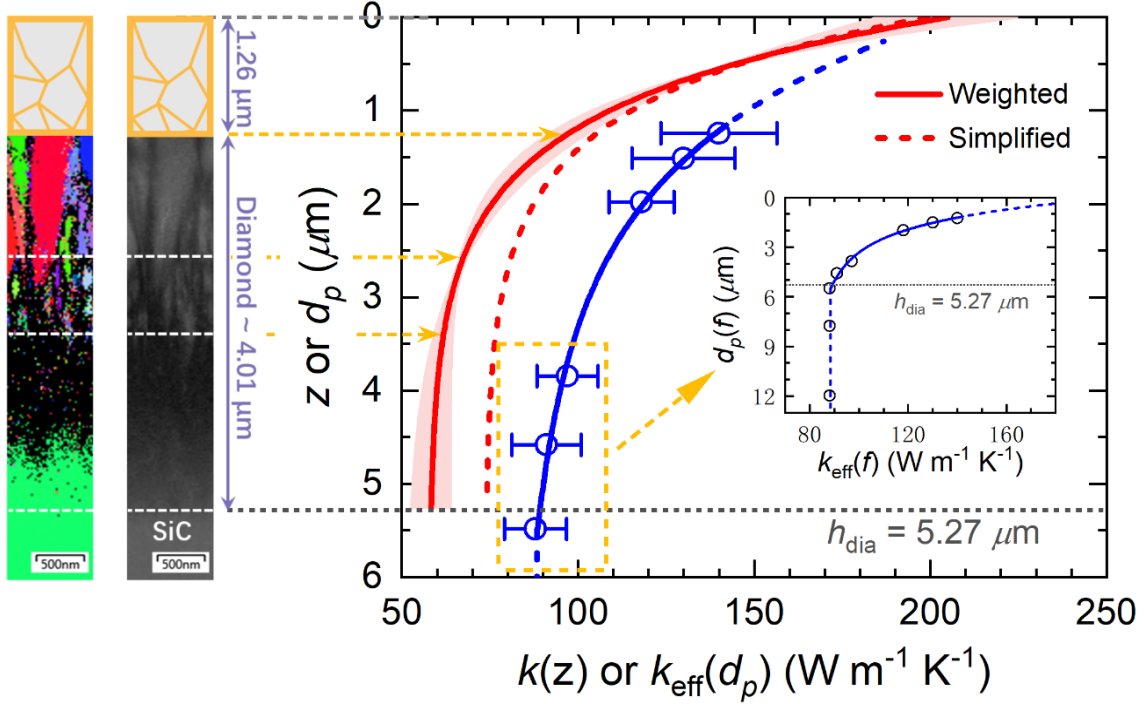

**Figure 4.** Depth-dependent thermal conductivity $k(z)$ reconstructed from the high-frequency SPS data. The solid red and pink dashed curves represent $k(z)$ obtained using the weighted harmonic mean model and the simplified harmonic mean model, respectively. Symbols show the measured effective conductivity $k_{\text{eff}}(f)$, while the solid and dashed blue curves correspond to the best-fit values. The shaded region indicates the uncertainty in the reconstructed profile. The monotonic increase in $k(z)$ corresponds to the grain-size evolution observed in the EBSD maps (left). The inset shows that the measured effective conductivity saturates at low frequencies where the thermal penetration depth exceeds the film thickness.

The shaded region in Fig. 4 represents the uncertainties in $k(z)$, which are substantially smaller than the individual uncertainties in $k_{\text{eff}}(f)$. This reduction arises from the low reduced chi-square obtained in the fitting process ($\chi^2 \approx 0.03$). Such a small $\chi^2$ reflects the presence of systematic, frequency-correlated uncertainties in input parameters—including the Al transducer thickness, laser spot size, and volumetric heat capacity—rather than uncorrelated random noise. Owing to the high repeatability of the SPS measurements, the six $k_{\text{eff}}(f)$ data points exhibit a smooth frequency dependence with minimal residual scatter. As a result, the combination of correlated parameter uncertainties and small residuals leads to $\chi^2 \ll 1$, indicating that the fit is well-constrained rather than overfitted. Notably, the uncertainty band narrows at specific depths (e.g., near $z \approx 0.5\ \mu\text{m}$ and $z \approx 2.7\ \mu\text{m}$) due to strong correlations among the profile parameters. Compensating variations in $k_0$, $k_1$, and $\alpha$ preserve the local value of $k(z)$, resulting in reduced uncertainty at these depths. A comprehensive uncertainty analysis is presented in Sec. S5 of the Supplementary Information, further confirming the robustness of the extracted $k(z)$.

To assess the sensitivity of our conclusions to the chosen functional form, we repeated the fitting using alternative models, including a sigmoidal profile and a two-region piecewise-linear model. In all cases, the reconstructed $k(z)$ exhibited the same monotonic increase, with details provided in Supplementary Information Sec. S6. Regardless of the model, the data unequivocally indicate a low-conductivity nucleation region, a rapid increase within the first $\sim 2\ \mu\text{m}$, and a near-surface limiting value. This demonstrates that the main physical conclusions are independent of the specific model chosen.

This behavior is further supported by features in the EBSD map. A pronounced grain-size reduction is observed between $z \approx 1.26\ \mu\text{m}$ and $z \approx 3.40\ \mu\text{m}$, coinciding with the region where $k(z)$ also decreases rapidly. Near the growth surface, the diamond film consists of well-faceted polycrystalline grains with characteristic lateral sizes on the order of a few hundred nanometers ($\sim 300$ nm). In contrast, toward the nucleation interface, the EBSD signal degrades markedly, indicating a nanocrystalline or partially amorphous transition zone with characteristic length scales well below the reliable EBSD resolution. Beyond $z \approx 3.40\ \mu\text{m}$, this transition zone corresponds to the depth where $k(z)$ reaches its minimum value of $\sim 60\ \text{W m}^{-1}\text{K}^{-1}$ and remains nearly constant. The strong correspondence between the grain-size evolution resolved by EBSD and the reconstructed thermal conductivity profile confirms that the depth dependence of $k(z)$ arises from the underlying grain size and phase evolution in the HFCVD diamond film.

In addition to grain boundary scattering, other microstructural factors may also contribute to the depth-dependent thermal conductivity of HFCVD diamond films. Point defects, variations in sp³/sp² bonding, non-diamond carbon phases, and extended defects such as dislocations are known to introduce additional phonon scattering, particularly near the nucleation interface.[25-28] While these effects are not independently resolved in the present measurements, they are expected to co-evolve with grain size during growth and to further suppress thermal transport in the nucleation region. Together, these mechanisms are consistent with the observed monotonic increase in $k(z)$ from the nucleation side toward the growth surface.

Interfacial thermal conductance $G$ between diamond and wide-bandgap semiconductors is also a critically studied parameter in high-power electronics. The findings from this study are contextualized within the broader literature, as summarized in Table I. A direct comparison is offered by Zhang et al.[29], who also investigated HFCVD diamond films on SiC using time-domain thermoreflectance (TDTR). While their reported diamond thermal conductivities (ranging from 21 to 241 W m⁻¹ K⁻¹) fall within a similar range to those observed here, their methodology could not resolve the depth-dependent conductivity profile, which is essential for explaining the order-of-magnitude variation in thermal properties with film thickness. Additionally, the limited modulation frequencies of TDTR likely hindered their ability to accurately extract the diamond/SiC interfacial thermal conductance ($G_2$), a key parameter central to our analysis.

The critical role of interlayers and processing in modulating $G_2$ is fundamentally linked to interfacial chemistry and structure. For instance, Yates et al.[30] demonstrated that a 5 nm SiN interlayer between MPCVD diamond and GaN increased the interfacial conductance fivefold, from 24 to 105 MW m⁻² K⁻¹. This improvement parallels the role of interlayers in our study, where they facilitate silicon diffusion and promote a more conductive, chemically bonded interface. The underlying mechanism is a transition from weak, contaminant-dominated interfaces to strong covalent bridging. This principle is further exemplified in diamond/SiC interfaces formed through surface active bonding (SAB), where post-annealing dramatically improves $G$. Values increase from $29\sim38$ MW m$^{-2}$ K$^{-1}$ for as-bonded interfaces[31, 32] to a maximum of 150 MW m$^{-2}$ K$^{-1}$ after high-temperature annealing,[31, 33] a result of defect reduction and enhanced chemical bonding.

Our results bridge and extend these previous findings. We not only quantify a competitive as-grown diamond/SiC interfacial conductance ($G_2 = 50 \pm 14$ MW m$^{-2}$ K$^{-1}$) for HFCVD films, but we also uniquely resolve the depth-dependent thermal conductivity within the diamond layer itself. Crucially, when contextualized with the literature, the data reveal two distinct regimes of interfacial quality: lower $G$ values ($\sim 20 - 50$ MW m$^{-2}$ K$^{-1}$) for as-deposited or poorly bonded interfaces, and higher $G$ values ($\sim 100 - 150$ MW m$^{-2}$ K$^{-1}$) for interfaces optimized via interlayers or annealing. This bimodal distribution is fundamentally explained by the quality of chemical bonding, transitioning from physically adsorbed, defective interfaces to coherent, chemically bonded ones. The ability of the SPS technique to simultaneously profile the diamond film and characterize the buried interface underscores a key advantage over methods like TDTR, providing a more complete thermophysical picture critical for the design and optimization of thermal management structures in high-power electronic devices.

Table I. Comparison of reported thermal properties of diamond-on-other-semiconductor structures.

| Material | Thickness ($\mu$m) | $k$ (W m$^{-1}$ K$^{-1}$) | $G$ (MW m$^{-2}$ K$^{-1}$) | Method | Data source |
|---|---|---|---|---|---|
| HFCVD diamond grown on 4H-SiC | 5.27 | 70-230 | 50 | SPS | This work |
| HFCVD diamond grown on Si | 0.63 | 241 | 18 | TDTR | Ref.[29] |
| HFCVD diamond grown on 3C-SiC | 7.8 | 21 | / | TDTR | Ref.[29] |
| MPCVD diamond grown on GaN without an interlayer | $1.0 \pm 0.14$ | $126^{+25}_{-22}$ | $24^{+8.2}_{-7.2}$ | TDTR | Ref.[30] |
| MPCVD diamond grown on GaN with SiN interlayer | $1.0 \pm 0.15$ | $132^{+22}_{-21}$ | $105^{+42}_{-19}$ | TDTR | Ref.[30] |
| 3C-SiC bonded with polycrystalline diamond | / | / | 38 (as-bonded) 150 (annealed) | TDTR | Ref.[31, 33] |
| 3C-SiC bonded with monocrystalline diamond | / | / | 34 (as-bonded) 111 (annealed) | TDTR | Ref.[31, 33] |
| 3C-SiC boned with monocrystalline diamond | / | / | 29 (as-bonded) | FD-PBD | Ref.[32] |

In conclusion, this study presents a comprehensive analysis of the depth-dependent thermal conductivity profile of an HFCVD diamond film on a SiC substrate, along with the quantification of key interfacial thermal resistances. Using the frequency-dependent SPS thermoreflectance technique, we observed that the effective thermal conductivity increases monotonically from the nucleation interface to the growth surface, a trend that correlates with grain coarsening observed via EBSD. This approach also enabled the simultaneous determination of the diamond/substrate interfacial conductance, a parameter typically inaccessible to higher-frequency techniques.

The methodology established herein provides a robust framework for characterizing anisotropic and heterogeneous material systems, which is crucial for optimizing thermal management in next-generation high-power electronic devices. While the exponential model offers a good fit to the experimental data, it assumes a uniform grain size evolution across the entire film, which may not hold in films with significant impurities or defects. Future work incorporating more complex models or higher-order terms could further refine our understanding of thermal conductivity in polycrystalline films and inform the design of advanced thermal management materials.

See the Supplementary Material for details of the optimization of HFCVD processes, derivation of the weighted harmonic mean model, input parameters, measured signals at all modulation frequencies, uncertainty analysis, evaluation of alternative models, and the MATLAB code used for the numerical calculations.

X.Q. acknowledges the funding support from National Key R&D program (2022YFA1203100). J.W. acknowledges support from the National Natural Science Foundation of China (NSFC) through Grant No. 52172037 and the Natural Science Foundation of Beijing through Grant No. L259007. P.J. acknowledges support from the National Natural Science Foundation of China (NSFC) through Grant No. 52376058.

## AUTHOR DECLARATIONS

### Conflict of Interest

The authors have no conflicts to disclose.

## DATA AVAILABILITY

The data that support the findings of this study are available from the corresponding authors upon reasonable request.